\documentclass[12pt]{amsart}
\usepackage{mathtools}

\DeclarePairedDelimiter\abs{\lvert}{\rvert}

\newcommand{\h}{Hochschild\ }
\newcommand{\hc}{Hochschild cochain\ }
\newcommand{\hg}{ho\-mot\-o\-py G-}
\newcommand{\coh}{co\-ho\-mol\-o\-gy}

\newcommand{\nc}{\mathbb{C}}
\newcommand{\nr}{\mathbb{R}}
\newcommand{\nz}{\mathbb{Z}}

\newcommand{\Ai}{$A_\infty$}
\newcommand{\BV}{\operatorname{B}V}
\newcommand{\End}[1]{\mathcal{E}nd_{#1}}
\newcommand{\F}{\underline{\mathcal {M}}(n)}
\newcommand{\Hom}{\operatorname{Hom}}
\newcommand{\id}{{\operatorname{id}}}
\newcommand{\IN}{{\operatorname{in}}}
\newcommand{\On}{\mathcal{O} (n)}
\newcommand{\tensor}{\otimes}
\newcommand{\X}[1]{\underline{\mathcal{M}}_{0,#1}}
\newcommand{\XF}{\underline{\mathcal {M}}(n) (X, \beta)}

\newtheorem{thm}{Theorem}
\newtheorem{lm}[thm]{Lemma}
\newtheorem{prop}[thm]{Proposition}
\newtheorem{claim}[thm]{Claim}
\newtheorem{crl}[thm]{Corollary}
\newtheorem{wish}[thm]{Wish}
\newtheorem{conj}{Conjecture} 

\theoremstyle{definition}

\newtheorem{df}{Definition}

\theoremstyle{remark}

\newtheorem{rem}{Remark} 
\newtheorem{ack}{Acknowledgment}

\begin{document}


\title{Homotopy G-algebras and moduli space operad}

\author
{Murray Gerstenhaber}
\address
{University of Pennsylvania}
\email{murray@math.upenn.edu}
\thanks{Research of the first author was supported in part by an NSA grant}

\author
{Alexander A. Voronov}
\email{voronov@math.upenn.edu}
\thanks{Research of the second author was supported in part by an NSF grant}


\begin{abstract}
This paper emphasizes the ubiquitous role of moduli spaces of
algebraic curves in associative algebra and algebraic topology. The
main results are: (1) the space of an operad with multiplication is a
\hg\ (i.e., homotopy graded Poisson) algebra; (2) the singular cochain
complex is naturally an operad; (3) the operad of decorated moduli
spaces acts naturally on the de Rham complex $\Omega^\bullet X$ of a
K\"{a}hler manifold $X$, thereby yielding the most general type of \hg
algebra structure on $\Omega^\bullet X$.
\end{abstract}

\maketitle

Recently, shortly after enunciation of a conjecture of Deligne \cite{d:lett},
the structure of a \hg\ (i.e., homotopy graded Poisson) algebra on the \hc
space $ V = C^\bullet(A,A) $ of an associative algebra was discovered
\cite{gj,gv}. It has also been pointed out in \cite{gj} that a similar
structure exists for the singular cochain complex $V = C^\bullet X$ of a
topological space, due to Baues \cite{baues}.

In this paper, we find a very general pattern which works for these two
examples: in both cases, $V$ has a natural structure of an operad. Together
with a multiplication, it yields the structure of a \hg algebra on $V$,
see Sections~1 and 2.

The rest of the paper is dedicated to the geometry of the conjecture, which, in
fact, assumed something more than mere algebraic structure:
\begin{conj}[Deligne]
\label{deligne}
The \hc complex has a natural structure of an
algebra over a chain operad of the little squares operad.
\end{conj}
In Section~4.2, we use the construction of Gromov-Witten invariants to
propose a way of proving a result analogous to the conjecture in the
case of a singular cochain complex: the singular chain operad of
decorated moduli spaces of genus zero algebraic curves acts naturally
on the de Rham complex $\Omega^\bullet X$ of a K\"{a}hler manifold
$X$, thereby yielding the most general type of \hg algebra structure
on $\Omega^\bullet X$.

\begin{ack}
We thank J.~Stasheff for helpful comments on an early version of the
manuscript. The second author would like to thank M.~Kontsevich and
Yu.~I. Manin for many useful discussions. He is also grateful to the
hospitality of the Max-Planck-Institut f\"{u}r Mathematik in Bonn,
where this paper was completed.
\end{ack}

\section{Homotopy G-algebra structure on an operad}

\subsection{The brace structure on an operad}
\label{brace structure}

Let $\{\On \; | \; n \ge 1\}$ be an operad of vector spaces. Such an
object relates to the collection of spaces $\Hom (V^{\otimes n} , V)$,
$n \ge 1$, of all $n$-ary operations on a vector space $V$ as much as
an abstract group relates to the group $\operatorname{GL} (V)$ of
linear transformations.
The structure of an operad on a collection of spaces $\On$ consists in
a number of operations $\gamma: \mathcal O (k) \otimes \mathcal O
(n_1) \otimes \dots \otimes \mathcal O (n_k) \to \mathcal O (n_1 +
\dots + n_k)$ satisfying certain natural associativity axioms. These
axioms can be read off from the \emph{endomorphism operad} $\End{V}
(n) := \Hom (V^{\otimes n} , V)$, where $\gamma$ is the substitution
of the values of $k$ operations in a $k$-nary operation as inputs. The
precise definition can also be read off from many recent papers on
operads, for example, \cite{gj,gk,ksv}.  In addition, one usually
assumes an action of the symmetric group $S_n$ on each $\On$ and a
unit element $\id \in \mathcal O (1)$ modeled after the action of
$S_n$ by permuting the inputs in $\Hom (V^{\otimes n} , V)$ and the
identity operator in $\Hom (V, V)$, respectively. We will not assume
an $S_n$ action here, i.e., work with non-$\Sigma$ operads.

Consider the graded vector space $\mathcal O = \bigoplus_n
\mathcal{O}(n)$, the sum of all components of the operad, now regarded as the graded components of $\mathcal O$.
Denote by
$\deg x$ the degree of an element $x \in \mathcal O$ and by $|x|$ the
degree in the {\it desuspension} $\mathcal O [1]$ of the graded vector
space $\mathcal O$, i.e., $\deg x = n$, $|x| = n-1$, whenever $x \in
\On$.  Define the following collection of multilinear operations, {\it
  braces} on $\mathcal O$:
\begin{equation}
\label{brace}
x\{ x_1, \dots, x_n\} := \sum
(-1)^\varepsilon \gamma
(x ; \id, \dots , \id, x_1, \id, \dots, \id, x_n, \id, \dots, \id)
\end{equation}
for $x, x_1, \dots , x_n \in \mathcal O$, where the summation runs
over all possible substitutions of $x_1, \dots, x_n$ into $x$ in the
prescribed order and $\varepsilon := \sum_{p=1}^{n} |x_{p}| i_{p} $,
$i_p$ being the total number of inputs of $\gamma (x; \id, \dots, \id,
x_1, \linebreak[0] \dots,\linebreak[1] \id, \linebreak[0] x_n,
\linebreak[0] \dots, \id)$ in front of $x_p$.  The braces $x\{ x_1,
\dots, x_n\}$ are homogeneous of degree $-n$ , i.e.,
\[
\deg{x\{ x_1, \dots, x_n\}} = \deg{x} + \deg {x_1} + \dots + \deg{x_n}
- n.
\]
We will
also assume the following conventions:
\[
x\{\} := x, \qquad x \circ y : = x\{y\}.
\]
\begin{rem}
The desuspension grading and sign are motivated by the example
$\End{V[1]}$ of the endomorphism operad of the desuspension $V[1]$,
where the vector space $V$ is regarded as a graded vector space
concentrated in degree 0.
In this case, the sign $(-1)^\varepsilon$ is picked up by rearranging
the sequence of letters $x\{ x_1, \dots, x_n\} (v_1, \dots, v_m)$,
where $v_1, \dots, v_m \in V[1]$, $m$ is such that $x\{ x_1, \dots,
x_n\} \in \Hom (V[1]^{\tensor m}, V[1])$, into the sequence
\[
\gamma (x ; v_1, \dots , v_{i_1}, x_1 (v_{i_1+1}, \dots), \dots,
v_{i_n}, x_n (v_{i_n+1}, \dots), \dots, v_m)
 \]
in accordance with the usual sign convention.
\end{rem}

One can immediately check the following identities:
\begin{multline}
\label{higher}
x\{x_1, \dots, x_m\} \{ y_{1}, \dots , y_{n}\} \\
= \sum_{0 \le i_1 \le \dots \le i_m \le n}  (-1)^\varepsilon
x\{ y_1, \dots,
y_{i_1}, x_1  \{ y_{i_1+1}, \dots , y_{j_1}\}, y_{j_1+1}, \dots,\\
y_{i_{m}}, x_m \{  y_{i_{m}+1} ,
\dots , y_{j_m}\}, y_{j_m+1}, \dots, y_n \} ,
\end{multline}
where $\varepsilon := \sum_{p=1}^{m}   |x_{p}| \sum_{q=1}^{i_p} |y_{q}|$, i.e.,
the sign is picked up by the $x_i$'s passing through the $y_j$'s in the
shuffle.

\begin{rem}
The identity for $m=n=1$ implies that the degree $-1$ bracket
\begin{equation}
\label{bracket}
[x,y] := x \circ y - (-1)^{|x| |y|} y \circ x
\end{equation}
defines the structure of a graded Lie algebra on $\mathcal O$.
\end{rem}

\begin{df}
A {\it brace algebra} is a graded vector space with a collection of braces
$x\{x_1, \dots, x_n\}$ of degree $-n$ satisfying the identities \eqref{higher}.
\end{df}
Thus we have made the following observation.
\begin{prop}
\label{braces}
For every operad $\mathcal O$ of vector spaces, the braces \eqref{brace} define the
natural structure of a brace algebra on the underlying graded vector space
$\mathcal O$.
\end{prop}

\subsection{Homotopy G-algebras}

A {\it multiplication on an operad} $\mathcal O$ is an element $m \in \mathcal O(2)$
such that $m \circ m =0$.

\begin{prop}
\label{d-dot}
\begin{enumerate}
\item
A multiplication on an operad is equivalent to a morphism $\mathcal{A}
s \to \mathcal O$ of operads, where $\mathcal{A} s$ is the associative
operad, cf. \cite{gk}.
\item If $V$ is an algebra over an operad with multiplication, then $V$ is
naturally an associative algebra.
\item The product
\end{enumerate}
\begin{equation}
\label{dot}
x \cdot y := (-1)^{\deg{x}} m \{x,y\}
\end{equation}
\begin{quote}
of degree $0$ and the differential
\end{quote}
\begin{equation}
\label{d}
dx := [m, x] = m \circ x - (-1)^{\abs{x}} x \circ m, \qquad d^2 =0, \quad \deg d =1,
\end{equation}
\begin{quote}
define the structure of a differential graded $($DG$)$ associative algebra on
$\mathcal O$.
\end{quote}
\end{prop}

Analogously, a {\it multiplication on a brace algebra} $V = \bigoplus_n V^n$ is
an element $m \in V^2$
 such that $m \circ m = 0$. It also provides $V$ with a DG algebra structure.

Amazingly, a multiplication generates a much richer algebraic structure on an
operad, as the following theorem implies. Before making the statement, we
should define the algebraic structure in question.

\begin{df}
A {\it \hg algebra\/} is a brace algebra $V = \bigoplus_n V^n$ provided with
a differential $d$ of degree one and a dot product $x  y$ of degree 0 making
$V$ into a DG associative algebra. The dot product must satisfy the following
compatibility identities:
\begin{equation}
\label{dist}
(x_1 \cdot x_2) \{ y_1, \dots, y_n\} = \sum_{k=0}^n (-1)^\varepsilon x_1 \{
y_1, \dots, y_k\} \cdot x_2 \{ y_{k+1}, \dots, y_n\},
\end{equation}
where $\varepsilon = |x_2| \sum_{p=1}^k |y_p|$, and
\begin{multline}
  \label{comm}
\begin{aligned}
&d ( x \{ x_1, \dots, x_{n+1}\} )
- (dx)\{x_1, \dots, x_{n+1}\} \\
&- (-1)^{|x|} \sum_{i=1}^{n+1} (-1)^{|x_1| + \dots + |x_{i-1}| }   x \{ x_1,
\dots, dx_i, \dots , x_{n+1}\}
\end{aligned}
\\
\begin{aligned}
\; = \; &  (-1)^{|x| |x_1|+1 } x_1  \cdot x \{ x_2, \dots, x_{n+1} \} \\
& + (-1)^{|x|}  \sum_{i=1}^n (-1)^{|x_1| + \dots + |x_{i-1}| } x \{ x_1, \dots
,
x_i \cdot x_{i+1}, \dots , x_{n+1} \}\\
& - 
 x \{ x_1, \dots, x_n \} \cdot
x_{n+1}
\end{aligned}
\end{multline}
\end{df}
\begin{rem}
1. Note that every \hg algebra is in particular a DG Lie algebra with respect
to the commutator \eqref{bracket},
which is a graded derivation of the dot product up to null-homotopy:
\begin{multline}
\label{deriv}
[x , yz] - [x,y] z - (-1)^{|x| (|y|+1)} y [x,z]
\\
 =  (-1)^{|x|+|y|+1}  (d ( x\{ y,z\} ) - (dx) \{  y, z\}
 \\
 - (-1)^{|x|} x \{dy,  z\}   - (-1)^{|x|+|y|} x \{y, d z\}) .
\end{multline}
Moreover, the multiplication is always {\it homotopy} graded commutative:
\begin{equation}
\label{hcomm}
xy - (-1)^{(|x|+1)(|y|+1)}  yx  = (-1)^{|x|} (d(x \circ y) - dx \circ y -
(-1)^{|x|}x \circ dy) .
\end{equation}

\noindent
2. Since a brace algebra $V$ with multiplication is a DG algebra, one
can define its \hc complex $C^\bullet (V, V)$ as usual, with the
differential
\[
D f := d \circ f -(-1)^{|f|} f \circ d + m \circ f - (-1)^{|f|} f \circ m
\]
and the cup product
\[
f_1 \cup f_2 := (-1)^{|f_1|+1} m \{ f_1, f_2 \},
\]
the degree $|f|$ meaning the desuspended total degree. (Usually, the
cup product has a different sign.) Then the relations \eqref{dist} and
\eqref{comm} mean that the correspondence
\begin{align*}
V & \to  C^\bullet (V, V), \\
x & \mapsto \sum_{n=0}^\infty x\{x_1, \dots, x_n\} ,
\end{align*}
the summation being in fact finite, is a morphism of DG algebras.
\end{rem}

\begin{thm}
\label{mainth}
A multiplication on an operad $\On$ defines the structure of a \hg algebra on
$\mathcal O = \oplus \On$. A multiplication on a brace algebra is equivalent to the
structure of a \hg algebra on it.
\end{thm}
\begin{proof}
The differential, the dot product and the braces have already been
defined. What remains is to check the compatibility identities. In
view of \eqref{dot} and \eqref{d}, they both are particular cases of
\eqref{higher}.
\end{proof}

\section{Applications: \h cochains and singular cochains}

\subsection{\h complex}

Applying this theorem to the \hc complex $C^\bullet (A, A)$, which is
at the same time the total space $\End{A}$ of the endomorphism operad
 $\End{A} (n) = C^n (A,A) = \Hom (A^{\tensor n}, \linebreak[0] A)$ as in Section \ref{brace structure} and whose multiplication cocycle $m (a,b) := a b $, $m \in C^2(A,A)$, is a multiplication on this operad, we obtain the following result conjectured by Deligne \cite{d:lett} and proved in \cite{gv,gj}.
\begin{crl}
The \h complex $C^\bullet (A, A)$ of an associative algebra $A$ has a natural
structure of \hg algebra.
\end{crl}

It is clear that the product obtained this way is the usual cup product
\begin{equation}
\label{dot-eq}
(x \cup y) (a_1, \dots , a_{k+l}) = x(a_1, \dots, a_k)
y (a_{k+1}, \dots , a_{k+l})
\end{equation}
altered by the sign $(-1)^{(|x|+1)(|y|+1)}$.  The bracket $[z,z]$
plays the role of a primary obstruction in deformation theory. It was
introduced by Gerstenhaber in \cite{gerst}. The higher braces for the
\h complex have been introduced in Kadeishvili and Getzler's works
\cite{kade, g:cartan}, where the brackets are used to define the \h
\coh\ of a homotopy associative algebra.

\subsection{G-algebras}

The structure inherited by the \h \coh\ was introduced in \cite{gerst} and has
been discovered in a number of places in mathematics and physics since then. A
{\it G-algebra\/} is a graded vector space $H$  with a dot product $xy$
defining the structure of a  graded commutative algebra and with a bracket
$[x,y]$ of degree $-1$ defining the structure of a graded Lie algebra, 
such that the
bracket with an element is a derivation of the
dot product:
\[
[x , yz] = [x,y] z + (-1)^{|x| (|y|+1)} y [x,z] .
\]
In other words, a G-algebra is a specific graded version of a Poisson algebra.
\begin{crl}
The dot product and the bracket
\[
    [x,y] := x \circ y - (-1)^{|x| |y|} y \circ x
\]
define the structure of a G-algebra on the \h \coh\ \linebreak[4]
$H^\bullet (A, \linebreak[0] A)$ of an associative algebra $A$.
\end{crl}

\begin{proof}
A simple computation shows that the identity \eqref{brace} yields the
Jacobi identity for the bracket. Equation \eqref{hcomm} implies that the
differential is a derivation of the bracket:
\begin{equation}
\label{d-br}
d[x, y] - [dx ,  y ] - (-1)^{|x|} [x , dy ] = 0.
\end{equation}
Therefore, even before passing to \coh, the Hochschild complex forms a DG Lie
algebra with respect to the bracket and a DG associative algebra with respect
to the dot product.

Thus, we will be through if we see that

\noindent
(i) the two operations take cocycles into cocycles and are independent of the
choice of representatives of \coh\ classes,

\noindent
and up to coboundaries,

\noindent
(ii) the dot product is  graded commutative and

\noindent
(iii) the bracket is a derivation of the dot product.

It easy to observe Fact (i) from the definition of the dot product and
the differential in any operad with multiplication (see \eqref{dot}
and \eqref{d}) and from the fact that $d$ is a derivation of the
bracket
\eqref{d-br}. Facts (ii) and (iii) have already been mentioned --- see
Equations~\eqref{deriv}
and \eqref{hcomm}.
\end{proof}

\subsection{Singular cochain complex}

Let $C^\bullet X$ be the singular cochain complex of a topological
space (or a simplicial set) $X$. For an $n$-simplex $\sigma: \Delta(n)
\to X$ with $\Delta (n)$ being the standard $n$-simplex, let $\sigma
(n_0, \dots , n_k)$ denote its face spanned by the vertices $n_0,
\dots, n_k$, where $i \mapsto n_i$ is an injective monotone function
$\Delta (k) \to \Delta (n)$. The singular cochain complex $C^\bullet
X$ has a natural operad structure, $S(n) = C^n X$, defined by the
compositions:
\[
  \gamma: S(k) \tensor S(n_1) \tensor \dots \tensor S(n_k) \to S(n_1 +
  \dots + n_k) ,
\]
\begin{align*}
\gamma(\varphi ; \varphi_1, \dots , \varphi_k ) (\sigma)
&: =  \varphi (\sigma (0, n_1, n_1 + n_2, \dots , n_1 + \dots + n_k))\\
& \qquad \varphi_1 (\sigma (0, 1, \dots, n_1 ))
\varphi_2 (\sigma (n_1, \dots, n_1 + n_2 )) \dots \\
& \qquad \varphi_k (\sigma (n_1 + \dots + n_{k-1}, \dots, n_1 + \dots + n_k )).
\end{align*}
This automatically yields the structure of a brace algebra on $C^\bullet X$,
according to Proposition~\ref{braces}.
Define the multiplication $m \in C^2 X$ as
\begin{equation*}
m(\sigma) := 1 \qquad \text{for any 2-simplex } \sigma .
\end{equation*}
Then Theorem~\ref{mainth} immediately recovers the following
statement, which is an interpretation of a result of Baues
\cite{baues}.

\begin{crl}
\label{baues}
The singular cochain complex $C^\bullet X$ of a topological space $X$ has a
natural structure of \hg algebra.
\end{crl}

The dot product determined by the multiplication $m$ as in \eqref{dot} is
nothing but the familiar cup product, up to the sign
$(-1)^{(|\varphi|+1)(|\psi|+1)}$,
\[
(\varphi \cup \psi ) (\sigma) = \sum_{k=0}^n \varphi (\sigma(0, \dots, k)) \psi ( \sigma
(k, \dots, n))
\]
and the differential determined by \eqref{d} is merely the familiar
coboundary operator, up to the sign $(-1)^{\abs{\varphi}+1}$,
\[
(d \varphi) (\sigma) = \sum_{k=0}^{n+1} (-1)^k \varphi (\sigma ( 0, 1,
\dots, \hat{k}, \dots, n+1)).
\]
Moreover, the brace $\varphi \{ \psi\}$ is the Steenrod operation $\varphi
{\cup}_1 \psi$ and higher braces are multilinear generalizations of it.

\section{Generalities}

\subsection{\Ai\ version}

Here we would like to consider \Ai\ generalizations of our results in
the spirit of \cite{gj}. An \Ai-{\it multiplication on an operad $\mathcal
O$ $($a brace algebra} $V$) is a formal sum $m = m_1 + m_2 +m_3 +
\dots$ with $m_n \in \On$ (or $V^n$), such that $m \circ m = 0$. In
this case \eqref{d} is also a differential, but not homogeneous, $d =
d_0 + d_1 + \dots$.  As above, an \Ai-multiplication on an operad
$\mathcal O$ defines
\begin{enumerate}
\item a morphism $\mathcal{A}_\infty \to \mathcal O$ of operads, where
$\mathcal{A}_\infty$ is the \Ai\ (homotopy associative) operad, see \cite{gk},
\item a natural structure of \Ai-algebra on each algebra over the operad $\mathcal
O$,
\item the structure of an \Ai-algebra on $\mathcal O$ itself with higher products
$M_n$ defined by the formula:
\begin{align*}
M_n(x_1, \dots , x_n) & := m\{x_1, \dots, x_n \} , \qquad \text{ for $n >
1$},\\
M_1 (x) & := dx := m \circ x - (-1)^{|x|} x \circ m.
\end{align*}
\end{enumerate}

For any \Ai-algebra $V$ the same formulas define the structure of an
\Ai-algebra on the Hochschild complex $C^\bullet (V, V)$, see Getzler
\cite{g:cartan}.

\begin{thm}
\label{A-inf}
An \Ai-multiplication on an operad $\On$ defines the following \Ai\ version of
a \hg algebra on $V= \mathcal O = \oplus \On$. It is a brace algebra $V$ and an
\Ai-algebra at the same time, such that  the correspondence
\begin{align*}
V & \to C^\bullet (V, V), \\
x & \mapsto \sum_{n=0}^\infty x\{ \dots \},
\end{align*}
is a morphism of \Ai-algebras.
\end{thm}

\subsection{Bar interpretation}

In this section we want to make a translation of the algebraic notions
introduced above into the dual language of bar constructions, following ideas
of Getzler-Jones' work \cite{gj}.
For a graded vector space $V = \bigoplus_n V_n$, let $V[-1] = sV$, $V[-1]_n
\linebreak[0] : = V_{n-1}$, be its {\it suspension} and
\[
\BV = \bigoplus_{n=0}^\infty (V[-1])^{\tensor n},
\]
the {\it bar coalgebra} with the usual coproduct
\begin{equation*}
\Delta [x_1 | \dots | x_n] = \sum_{i=0}^n [x_1| \dots |x_i] \tensor [x_{i+1} |
\dots |x_n],
\end{equation*}
$[x_1 | \dots | x_k]$ denoting an element of $V[-1]^{\tensor k} \subset \BV$.

We call a product $\BV \tensor \BV \xrightarrow{\cup} \BV$ {\it left nonincreasing} if
$\deg(x \cup y) \ge \deg x$, where by definition $\deg [x_1 | \dots | x_n] =
n$.
\begin{lm}
\label{noninc}
The structure of a brace algebra on a graded vector space $V$ is equivalent to
the structure of a bialgebra on the bar coalgebra $\BV$ defined by a left
nonincreasing product.
\end{lm}

\begin{proof}
A product $\cup$ on $\BV$ compatible with the coproduct $\Delta$ and satisfying
$\deg(x \cup y) \ge \deg x$ determines the braces uniquely by the formula
\begin{multline*}
[x_1| \dots | x_m] \cup [y_1| \dots |y_n] \\
= \sum (-1)^\varepsilon [y_1 | \dots | y_{i_1}|
x_1\{y_{i_1+1}, \dots \} | \dots | y_{i_m}|
x_m\{y_{i_m+1}, \dots \} | \dots | y_{n}] ,
\end{multline*}
where the sign is the same as in \eqref{higher}.
The associativity of the product is then equivalent to the relations
\eqref{higher}.
\end{proof}

Let $V$ be a brace algebra, $\BV$ the corresponding bar bialgebra. A DG-{\it
bialgebra} is a bialgebra with a degree $-1$ differential which is
simultaneously a derivation and a coderivation.

\begin{lm}
\label{DG-bi}
An \Ai-multiplication on a brace algebra $V$ is equivalent to the structure of
a DG-bi\-al\-ge\-bra on the bar bialgebra $\BV$.
\end{lm}

\begin{proof}
An \Ai-multiplication $m$ on $V$ is equivalent to a codifferential $\delta$ on
$\BV$, as has been well-known since Stasheff \cite{jim}. That $\delta$ is a
derivation is equivalent to the compatibility condition of Theorem~\ref{A-inf}.
\end{proof}

\section{Topological and mirror reflections}

\subsection{Moduli spaces and little squares}
\label{config}

As predicted by Deligne \cite{d:lett}, the structure of a \hg algebra
on the \h complex arises from an action of a chain complex of the
little squares operad. The following combinatorial version of this
statement was proved by Getzler and Jones. Consider Fox-Neuwirth's
cellular partition of the configuration spaces $F(n, \nr^2)$ of $n$
distinct points in $\nr^2$: cells are labeled by ordered partitions
of the set $\{1, \dots, n\}$ into subsets with orderings within each
subset. This reflects grouping points lying on common vertical lines
on the plane and ordering the points lexicographically. For each $n$,
take the quotient cell complex $K_\bullet \mathcal M (n)$ by the action of
translations $\nr^2$ and dilations $\nr^*_+$. These quotient spaces do
not form an operad, but one can glue lower $K_\bullet \mathcal M (n)$'s to
the boundaries of higher $K_\bullet \mathcal M (n)$'s to form a cellular
operad $K_\bullet \F$. The resulting space $\F$ is a circle bundle
over the real compactification $\X{n}$ of the moduli space $\mathcal
M_{0,n}$ of $n$-punctured curves of genus zero, see
\cite{as,gj,ksv,kon}. The space $\F$ can be also interpreted as a
``decorated'' moduli space, see next section. Cells in this cellular
operad $K_\bullet \F$ are enumerated by pairs $(T, p)$, where $T$ is a
tree with $n$ initial vertices and one terminal vertex, labeling a
component of the boundary of $\F$, and $p$ is a partition, as above,
of the set $\IN (v)$ of incoming vertices for each vertex $v$ of the
tree $ T$.

In \cite{gj}, it is shown that a complex $V$ is an \Ai\ \hg algebra, iff it is
an algebra over the operad $K_\bullet \F$ satisfying the following condition.
The structure mappings
\[
K_\bullet \F \to \Hom (V^{\tensor n}, V),
\]
of the algebra $V$ over the operad $K_\bullet \F$ send all cells in $K_\bullet
\mathcal{M} (n)$ to zero, except cells of two kinds:
\begin{enumerate}
\item $(\delta_n, ((i_1), (i_2, \dots, i_n))$, where $\delta_n$ is the corolla,
the tree with one root and $n$ edges, connecting it to the remaining $n$
vertices, corresponding to the configuration where the points $i_2, \dots, i_n$
sit on a vertical line, the $i_k$th point being below the $i_{k+1}$st, and the
$i_1$st point is in the half-plane to the left of the line;
\item $(\delta_n, ((i_1, i_2, \dots, i_n))$, corresponding to the configuration
where all the points sit on a single vertical line, the $i_k$th point being
below the $i_{k+1}$st.
\end{enumerate}
Cells of the first kind map to the braces $x_{i_1}\{ x_{i_2}, \dots , x_{i_n}
\}$, $n \ge 1$, and cells of the second kind map to the \Ai\ products $M_n
(x_{i_1}, \dots, x_{i_n})$, $n > 1$.

Thus, the conditions of Theorem~\ref{A-inf} and, in particular, when $M_n = 0$
for $n > 2$ the relations \eqref{dist} and \eqref{comm}, follow from the
combinatorial structure of the cell complex.

This construction of Getzler and Jones is essentially combinatorial and the
question of a natural topological construction, perhaps, similar to those which
come from quantum field theory (cf. \cite{ksv}), where algebraic operations are
obtained by integration over cycles, remains a mystery. At least in known
examples, we anticipate that other cells give rise to nonzero multilinear
operations.

\begin{wish}
\label{w}
In all the examples, where a \hg algebra structure occurs, e.g., the \h complex
or the singular cochain complex, the structure extends nontrivially to a
natural structure of an algebra over the operad $K_\bullet \F$.
\end{wish}

This operad is closer to the \hg operad in the sense of
Ginzburg-Kapranov \cite{gk}. The latter operad is the operad cobar
construction for the Koszul dual to the Gerstenhaber operad, which is
the homology operad $H_\bullet \F$, see \cite{C1,gj}.

\subsection{The construction}

Here we are going to sketch how to make Wish~\ref{w} come true in case of the
\hg algebra $C^\bullet X$. Our construction is a real version of Kontsevich's
construction of Gromov-Witten invariants in \cite{k:mirror,konm}. The
difference is that we replace the moduli spaces $\overline{\mathcal M}_{0,n+1}$,
which are compact complex manifolds, with the spaces $\F$, which are circle
bundles over the real compactifications $\X{n+1}$, which are compact real
manifolds with corners.

Let $(X, \omega)$ be a compact manifold with a sufficiently positive K\"{a}hler
 form $\omega$. We will replace singular cochains $C^\bullet X$ with smooth
forms
\[
V = \Omega^\bullet X .
\]
We want to define the natural structure of an algebra over the singular chain
operad $C_\bullet \F$ on $V$. In particular, this will yield the structure of
an algebra over the cellular chain operad $K_\bullet \F$ on $V$, giving a
solution of Wish~\ref{w}.

Let $\XF$ be the moduli space of {\em stable
holomorphic maps} $(C; p_1, \dots, p_{n+1}; \linebreak[0] \tau_1, \dots,
\linebreak[0] \tau_m,
\tau_\infty; \phi )$ from a (degenerated) curve $C$ of ge\-nus 0 to $X$:
\[
\phi: C \to X
\]
mapping the fundamental class of $C$ to a given homology class $\beta \in
H_2(X, \nz)$. Here the curve $C$ has $n+1$ punctures $p_1, \dots, p_{n+1}$ and
all the singularities of $C$ must be $m$ double points. For each $i$, $ 1 \le i
\le m$, $\tau_i$ is the choice of a tangent direction at the $i$th double point
to the irreducible component that is farther away from the ``root'', i.e., from
the component of $C$ containing the puncture $\infty := p_{n+1}$.
$\tau_\infty$ is a tangent direction at $\infty$. The stability of a map is
understood in the sense of Kontsevich \cite{k:mirror,konm}: each irreducible
component of $C$ contracted to a point by $\phi$ must be stable, i.e., admit no
infinitesimal automorphisms. Because of a finite group of automorphisms, the
moduli space $\XF$ is only a compact stack. In some cases, e.g., when $X$ is a
homogeneous space, it is a smooth stack with corners. Let us assume it is one.

Notice that the configuration space $\F$ considered in Section~\ref{config} is
the same as the moduli space of data $(C; p_1, \dots, p_{n+1}; \tau_1, \dots,
\tau_m,
\tau_\infty)$ as above, except that all components of $C$ must be stable, cf.
\cite{gj,ksv}. The operad composition is given by attaching the $\infty$
punctures on curves to the other punctures on another curve, remembering the
tangent direction at each new double point. Let
\[
\pi: \XF \to \F
\]
denote the forgetful map of the space $\XF$ of maps to the space $\F$ of
curves.

There is a universal (evaluation) map
\[
\Phi: \mathcal{C} \to X
\]
from the universal curve over $\XF$ to the manifold $X$. The natural projection
$ \mathcal{C} \to \XF$ admits $n+1$ canonical sections $s_1, \dots, s_{n+1}$,
sending a point of the moduli to the $i$th puncture on the universal curve.

Now we are ready to define the structure of an algebra over the operad
$C_\bullet \F$ on $V = \Omega^\bullet X$. Let $(\varphi, \psi) := \int_X
\varphi \wedge \psi$ be the Poincar\'{e}
pairing on $V$. Leaving aside problems with pairings and duals for infinite
dimensional vector spaces and replacing singular cochains with differential
forms once again, it suffices to construct mappings
\begin{equation}
\label{GW}
f_n : V^{\tensor n+1} \to \Omega^\bullet \F
\end{equation}
which will define the structure of an algebra over the operad
\[
C_\bullet \F \to \Hom(V^{\tensor n}, V)
\]
after dualizing $V$ with the help of the Poincar\'{e} pairing. (Honestly
speaking, we would have to replace singular chains with currents then). We
define the mapping \eqref{GW}
by the formula
{\small
\[
f_n (\varphi_1, \dots , \varphi_{n+1}) := \sum_{\beta \in H_2 (X, \nz)} \exp
(-\int_\beta \omega) \; \pi_* (s_1^* \Phi^* \varphi_1 \wedge \dots \wedge
s_{n+1}^* \Phi^* \varphi_{n+1}),
\]
}where $\Phi^*$ and $s_i^*$'s denote pullbacks and $\pi_*$ a push-forward
(fiberwise integration). In fact, because of the summation over the lattice
$H_2(X, \nz)$, we have to replace the ground field $\nc$ with formal power
series in $\beta \in H_2(X, \nz)$.

\begin{claim}
The maps \eqref{GW} define a morphism of operads, that is, the structure of an
algebra over the chain operad $C_\bullet \F$ on the de Rham complex $V =
\Omega^\bullet X$.
\end{claim}

As soon as some hard problems with the construction, such as the smoothness of
the stack of stable maps, are solved, the verification of the operad properties
of this claim is automatic. So, we postpone the proof of it until better times.


\begin{thebibliography}{10}

\bibitem{as} S.~Axelrod and I.~M. Singer, \emph{Chern-{S}imons
perturbation theory II}, J. Diff. Geom. {\bf 39} (1994), 173--213,
  hep-th/9304087.

\bibitem{baues}
J.~H. Baues, {\em The double bar and cobar constructions}, Compos. Math. {\bf
  43} (1981), 331--341.

\bibitem{C1}
F.~R. Cohen, {\em The homology of {$\mathcal{C}_{n+1}$}-spaces, {$n\ge0$}}, The
  homology of iterated loop spaces, Lecture Notes in Math., vol. 533,
  Springer-Verlag, 1976, pp.~207--351.

\bibitem{d:lett}
P.~Deligne, {\em Letter to {S}tasheff, {G}erstenhaber, {M}ay, {S}chechtman,
  {D}rinfeld}, May 17, 1993.

\bibitem{gerst}
M.~Gerstenhaber, {\em The cohomology structure of an associative ring}, Ann. of
  Math. {\bf 78} (1963), 267--288.

\bibitem{gv} M.~Gerstenhaber and A.~A. Voronov, {\em Higher operations
  on {H}ochschild complex}, Function.\ Anal.\ Appl. \textbf{29}
  (1995), no.\ 1, 1--6.

\bibitem{g:cartan}
E.~Getzler, {\em Cartan homotopy formulas and the {G}auss-{M}anin connection in
  cyclic homology}, Israel Math. Conf. Proc. {\bf 7} (1993), 65--78.

\bibitem{gj}
E.~Getzler and J.~D.~S. Jones, {\em Operads, homotopy algebra and iterated
  integrals for double loop spaces}, Preprint, Department of Mathematics, MIT,
  March 1994, hep-th/9403055.

\bibitem{gk}
V.~Ginzburg and M.~Kapranov, {\em Koszul duality for operads}, Duke Math. J.
{\bf 76} (1994), 203--272.

\bibitem{kade} T.~V. Kadeishvili, \emph{The structure of the
$A(\infty)$-algebra, and the Hochschild and Harrison cohomologies},
  Trudy Tbiliss. Mat. Inst. Razmadze Akad. Nauk Gruzin. SSR
  \textbf{91} (1988), 19--27.
  
\bibitem{ksv} T.~Kimura, J.~Stasheff, and A.~A. Voronov, {\em On
  operad structures of moduli spaces and string theory},
  Commun.\ Math.\ Phys. \textbf{171} (1995), 1--25, hep-th/9307114.

\bibitem{kon} M.~Kontsevich, {\em Feynman diagrams and low-dimensional
  topology}, First European Congress of Mathematics, Vol.\ II (Paris,
  1992), 97--121.  Progr.\ Math., 120, Birkh\"{a}user Verlag, Basel,
  1994.

\bibitem{k:mirror} \bysame, {\em Enumeration of rational curves via
  torus actions}, The moduli space of curves (Texel Island, 1994),
  335--368. Progr.\ Math., 129, Birkh\"{a}user Boston, Inc., Boston,
  MA, 1995, hep-th/9405035.

\bibitem{konm} M.~Kontsevich and Yu. Manin, {\em Gromov-Witten
  classes, quantum cohomology, and enumerative geometry},
  Comm.\ Math.\ Phys.\ \textbf{164} (1994), no.\ 3, 525--562,
  hep-th/9402147.

\bibitem{jim} J.~D. Stasheff, \emph{On the homotopy associativity of
{H}-spaces II}, Trans.\ Amer.\ Math.\ Soc. {\bf 108} (1963),
  293--312.

\end{thebibliography}

\makeatletter \renewcommand{\@biblabel}[1]{\hfill#1.}\makeatother

\end{document}